\documentclass[pre,onecolumn,12pt]{revtex4}

\usepackage{graphicx}
\usepackage[caption=false]{subfig}
\usepackage{amsmath}
\usepackage{amssymb}
\usepackage{amsthm}
\usepackage{bm}
\usepackage{cancel}
\usepackage[normalem]{ulem}
\usepackage{xfrac}

\usepackage{color}

\usepackage{hyperref}
\usepackage{cleveref}

\usepackage[section]{placeins}
\usepackage{comment}

\newcommand{\short}{GaAs/Al$_{0.45}$Ga$_{0.55}$As}
\newcommand{\tall}{GaAs/Al$_{0.7}$Ga$_{0.3}$As}
\begin{document}

\graphicspath{{images/}}

\title{Parameter dependence of high-frequency nonlinear oscillations and intrinsic chaos in short GaAs/(Al,Ga)As superlattices}

\author{Jonathan Essen${}^\star$${}^\ddagger$\footnote{Corresponding author, email: essen@physics.ucsb.edu}, Miguel Ruiz-Garcia${}^\dagger$, Ian Jenkins${}^\star$${}^\ddagger$, Manuel Carretero${}^\dagger$,\\ Luis Bonilla${}^\dagger$ and Bj\"orn Birnir${}^\ddagger$${}^\circ$}
\address{${}^\dagger$Gregorio Mill\'an Institute for Fluid Dynamics, Nanoscience and Industrial Mathematics, and Department of Materials Science and Engineering and Chemical Engineering, Universidad Carlos III de Madrid,
Avenida de la Universidad 30, 28911 Legan\'es, Spain}
\address{${}^\star$Department of Physics, University of California, Santa Barbara, 93106, United States}
\address{${}^\ddagger$Department of Mathematics and CNLS, University of California, Santa Barbara, 93106, United States}
\address{${}^\circ$University of Iceland, 107 Reykjav\'ik, Iceland}

\begin{abstract}
We explore the design parameter space of short (5--25 period), n-doped, Ga/(Al,Ga)As semiconductor superlattices (SSLs) in the sequential resonant tunneling regime. We consider SSLs at cool (77K) and warm (295K) temperatures, simulating the electronic response to variations in (a) the number of SSL periods, (b) the contact conductivity, and (c) the strength of disorder (aperiodicities). Our analysis shows that the chaotic dynamical phases exist on a number of sub-manifolds of codimension zero within the design parameter space. This result provides an encouraging guide towards the experimental observation of high-frequency intrinsic dynamical chaos in shorter SSLs.
\end{abstract}


\maketitle

%
%
%
%
%
%




\section{Introduction}
Chaotic oscillations in n-doped, weakly-coupled semiconductor superlattices (SSLs) have generated interest for many years. Spontaneous oscillations, quasiperiodic orbits, and chaos have already been observed experimentally at very low temperatures \cite{Leo92, Ka97,Wu99} and at room temperature \cite{li2013fast, kanter10, reid09, huang2012experimental,huang2013spontaneous} in 50-period SSLs. 
The sequential resonant tunneling (SRT) model of Bonilla et.\ al.\ (see reviews in \cite{bonilla2009nonlinear, wacker2002semiconductor}) captures the essential physics of tunneling transport in SSLs. The model contains nonlinearities arising from feedback between resonant tunneling through the barriers and the self-consistent electric field of the mobile carriers. Simulations of the 50-period SSLs by Alvaro et.\ al.\ \cite{alvaro2014noise} demonstrated extreme sensitivity to weak stochastic perturbations of the local electric fields and the bias voltage, which provided a qualitative description of the experimental results. However, the unperturbed dynamics of the 50-period SRT model contained only a period doubling bifurcation, rather than fully-developed chaos.
Recently, it was observed that shorter (10-period) SSLs, support chaotic oscillations on much faster timescales \cite{garcia2017}. In contrast with the 50-period simulations, the chaos in shorter SSLs exhibited a complete period-doubling cascade.

One of the practical applications of chaotic oscillations in SSLs is the secure generation of random bit sequences. With faster chaotic oscillations, shorter SSLs would achieve a higher random bit rate. 
In this work, we provide aid to the experimental search for period-doubling cascades in shorter SSLs by mapping out the response of the SRT model to variations of the basic design parameters.

We simulate the SRT theory, which describes electronic transport in SSLs in the weakly-coupled, self-consistent regime. Two different time scales are taken into account in this description: The inter-site tunneling and inter-subband relaxation processes occur on much shorter timescales than the dielectric relaxation processes \cite{bonilla2000microscopic}. Therefore, the long timescale dynamics of semiconductor lasers \cite{scia15} and superlattices \cite{bonilla2005nonlinear,alvaro2014noise} are typically modeled using semiclassical equations, while the short timescale processes are treated through the addition of stochastic terms to the dynamical equations. Nonlinearities enter the model via the the self-consistent electron-electron Coulomb interaction, which bends the conduction band of the SSL, modifying the inter-subband tunneling rates by casting the energy levels of adjacent wells into or out of resonance \cite{bonilla2000microscopic}.

Over some intervals of the bias voltage, the total current $J(t)$ through the SSL is a monotonically increasing function of the bias voltage. At higher bias voltages, $J(t)$ suddenly changes to a time-dependent, oscillatory function, which undergoes a series of transitions, leading to chaotic behavior. We summarize the behavior of $J(t)$ below:
\paragraph*{Bistability:} The earliest signal of the oscillatory behavior is a bistable response of $J(t)$ to slow variations in $V_{\text{bias}}$. This behavior is observable only at sufficiently low temperatures \cite{bonilla2006voltage,guido2006hopf,luo1999controllable}.
Generically, bistable behavior is found at voltages just below those of the Hopf bifurcation described next.
\paragraph*{Supercritical Hopf Bifurcation:} As the bias voltage is increased, $J(t)$ undergoes a supercritical Hopf bifurcation. The fixed point becomes unstable, and $J(t)$ becomes attracted to a closed periodic orbit.
In this regime, the SSL acts as a GHz oscillator with a discrete power spectrum involving the frequencies $f_n = n/T,\ n=1,2,3,...$, where the \emph{fundamental period} $T$ is the period of the lowest-frequency oscillation present. The superharmonics $n>1$ arise due to the nonlinearities of the SRT model. Since $T$ varies smoothly with the bias voltage, the oscillator is also \emph{tunable}.
\paragraph*{Period Doubling Bifurcation:}
The periodic orbit described above is topologically equivalent to a circle in phase space. The Poincar\'e map of this trajectory consists of a single point, called a \emph{one-cycle}. Increasing the bias voltage further, one-cycles of the Poincar\'e map transition to two-cycles, i.e.\ two points, meaning that the orbit circles twice before it closes onto itself. The fundamental period of the oscillator is doubled, $T\to2T$, and the fundamental frequency is cut in half: $f_1\to f_1/2$. A new peak appears in the power spectrum spectrum at half the fundamental frequency, and the number of superharmonics doubles. Following a period doubling bifurcation, it is possible that the reverse (period-halving) bifurcation may occur. We refer to the regions between these bifurcations as \emph{period doubling bubbles}.
An application of period doubling, due to the subharmonic peak, is the generation of \emph{squeezed states} \cite{Gr87}, which have applications in the area of noise reduction.
\paragraph*{Period Doubling Cascade:} 
An infinite number of period doublings is possible within a finite voltage interval. The invariant phase space structures transition from a smooth compact manifolds (periodic orbits of high periods) to irregular sets called \emph{strange attractors}. The Poincar\'e map takes on a fractal structure.

Transport in SSLs can take place through two possible channels: Quantum tunneling between the $\Gamma$ valleys of adjacent wells, or phonon-assisted transport through the $X$ valley of the barriers ($\Gamma$-$X$ transfer). Chaotic oscillations in SSLs are only possible when tunneling transport dominates  over diffusive transport \cite{huang2012experimental, huang2013spontaneous}. The phonon-assisted transport may be suppressed by (a) lowering the temperature of the SSL, (b) reducing the level of doping (and hence the Fermi energy), or (c) adding Aluminum to the barriers. For GaAs/AlAs SSLs, the $X$-valley of the AlAs barriers is only 110~meV higher than the lowest subband of the 4~nm GaAs wells considered in this paper. Adding Aluminum to the barriers has the effect of lowering the $\Gamma$-minimum and increasing the $X$-minimum. For Al$_x$Ga$_{1-x}$As SSLs with Aluminum concentration $x=0.45$, the $\Gamma$- and $X$-minima are both 337meV above the lowest subband of the GaAs well. Therefore at room temperature, the $x=0.45$ SSLs supress the phonon-assisted transport by a factor of about $1.6\times 10^{-4}$ compared with the $x=0$ SSLs \cite{huang2012experimental}. In this work, we fix the doping density to $N_D=6\times10^{10}\textrm{ cm}^{-2}$ 
and simulate two scenarios: \tall{} SSLs at 77K and \short{} SSLs at 295K.

With an eye towards development of fast, electronic true random number generators, we investigate the response of the chaotic signal to variations of the design parameters of these systems: The number of periods, the contact conductivity, and the strength of the \emph{disorder} (aperiodicity) of the SSL. The outline of our paper is as follows:
In Section~\ref{sec:model}, we review the SRT model.  In Section~\ref{sec:method}, we describe the numerical methods. In Section~\ref{sec:results}, we present the results of our simulations. A discussion of our results is given in Section~\ref{sec:conclusion}.


\section{Model}\label{sec:model}
Many phenomena are captured by SRT model of nonlinear charge transport in SSLs \cite{bonilla2005nonlinear,bonilla2002,bonilla2009nonlinear, wacker2002semiconductor}. Consider a weakly coupled superlattice having $N$ identical periods of length $l$ and total length ${ L=N l }$ subject to a dc bias voltage $V_{\text{bias}}$. The time evolution of the average electric field of SSL period $i$, $F_i$, and the total current density, $J(t)$, are coupled together by Ampere's law
\begin{equation}\label{eq:ST1}
J(t) = \epsilon \frac{dF_i}{dt}+J_{i\to i+1},
\end{equation}
with the voltage bias constraint
\begin{equation}
    \label{eq:bias}
		\sum_{i=1}^{N}F_i=\frac{V_{\text{bias}}}{ l }.
\end{equation}
Fluctuations of $F_i$ away from its average value ${F_{\text{avg}}=eV_{\text{bias}}/L}$ arise from the inter-site tunneling current $J_{i\to i+1}$, which appears in equation~\eqref{eq:ST1}. A microscopic derivation of $J_{i\to i+1}$ produces the result \cite{bonilla2000microscopic,bonilla2002}
\begin{equation}
\label{eq:tunnel}
J_{i\rightarrow i+1}=\frac{e n_i}{l} v^{(f)}(F_i)-J_{i\rightarrow i+1}^{-}(F_i,n_{i+1},T),
\end{equation}
in which $n_i$ is the electron sheet density at site $i$, $-e<0$ is the electron charge and $T$ is the lattice temperature. The forward velocity, $v^{(f)}(F_i)$, which is modeled as a Lorentzian distribution, is peaked at resonant values of $F_i$, where the lowest energy level at site $i$ are aligned with one of the levels at site $i+1$. The backward tunneling current is given by
\begin{equation}
\label{eq:backwards}
\begin{split}
&J_{i\rightarrow i+1}^{-}(F_i,n_{i+1},T)= \\
&\ \ \frac{em^* k_B T}{\pi \hbar^2 l}v^{(f)}(F_i) \ln \left[1+ e^{-\frac{eF_i l}{k_B T}}\left( e^{\frac{\pi \hbar^2 n_{i+1}}{m^* k_B T}} -1\right)  \right]\!,
\end{split}
\end{equation}
where the reference value of the effective electron mass in Al$_x$Ga$_{1-x}$As is $m^*=(0.063 +0.083x) m_e$, and $k_B$ is the Boltzmann constant. The $n_i$ are determined self-consistently from the discrete Poisson equation,
\begin{equation}
n_i=N_D +\frac{\epsilon}{e}(F_i-F_{i-1}),
\end{equation}
where $N_D$ is the doping sheet density and $\epsilon$ is the average permittivity. The field variables $F_i$ are constrained by boundary conditions at $i=0$ and $i=N$ that represent Ohmic contacts with the electrical leads
\begin{equation}\label{eq:STN}
J_{0 \rightarrow 1}=\sigma_0 F_0, \quad J_{N \rightarrow N+1} =\sigma_0 \frac{n_{N}}{N_D} F_N,
\end{equation}
where $\sigma_0$ is the contact conductivity. Shot and thermal noise can be added as indicated in \cite{alvaro2014noise,bon2016jmi}.

\begin{table}[t]
	\centering
	\begin{ruledtabular}
	\begin{tabular}{llllll}
		$N_D$ (cm${}^{-2}$)          & $d$ (nm)& $w$ (nm)&
		$s$ ($\mu$m) \\ \hline
		$6 \times 10^{10}$ & $4$ & $7$ &
		$60$
	\end{tabular}
    \vspace{1ex}
	\begin{tabular}{lllll}
		V$_{barr}$ (meV) & $T$ (K) & E$_1$ (meV) & E$_2$ (meV) & E$_3$ (meV)\\ \hline
		$600$           & $77$      & $53$     & $207$          & $440$ \\
		$388$           & $295$    & $45$     & $173$          & $346$
	\end{tabular}
	\end{ruledtabular}
	\caption{(Top) The design parameters of the superlattice. (Bottom) Values of the potential barrier and energy levels for GaAs/Al$_{0.7}$Ga$_{0.3}$As and GaAs/Al$_{0.45}$Ga$_{0.55}$As superlattices, first and second row, respectively.}
	\label{tab:params}
\end{table}

Table~\ref{tab:params} gives the numerical values of the parameters used in the simulations. The GaAs/Al$_{0.45}$Ga$_{0.55}$As configuration corresponds with the configuration used in recent experiments \cite{li2013fast,huang2012experimental,Yin16}. The rest of the parameters are as follows: $A=s^2$ is the transversal area of the superlattice, $d$ and $w$ are the barrier and well widths, and $l=d+w$ is the SSL period. The contact conductivity $\sigma_0$ is a linear approximation of the behavior of $J_{0\rightarrow 1}$, which depends on the structure of the emitter. We treat $\sigma_0$ as an empiricial parameter and investigate the response of the SRT model as it is varied. Some representative values have been chosen in order to reproduce the experimental results produced by Huang \emph{et al.} with $N=50$: $\sigma_0=0.783$ A/Vm for $V_{barr}=388$ meV ($x=0.45$) and $\sigma_0=0.06$ A/Vm for $V_{barr}=600$ meV ($x=0.7$), where $V_{barr}$ is the height of the barrier \cite{alvaro2014noise,huang2012experimental}.

\subsection{Noise} 
To model the unavoidable fluctuations in the bias voltage, as well as the short-timescale processes in the electronic dynamics, stochastic terms \cite{alvaro2014noise} are introduced into equations \eqref{eq:ST1}--\eqref{eq:STN}. To account for the noise in the bias voltage, equation \eqref{eq:bias} is modified to
\begin{equation}
  \label{eq:bias_noise}
\sum_{i=1}^{N}F_i=\frac{V_{\text{bias}} + \eta(t)}{l},
\end{equation}
where $\eta(t)$ is taken to be a Gaussian random variable with standard deviation $\sigma_\eta$. To account for the short-timescale processes at each site of the SSL, equation \eqref{eq:ST1} is modified to include \emph{shot noise} in the local tunneling current
\begin{equation}
  \label{eq:ampere_noise}
\epsilon \frac{dF_i}{dt}+J_{i\to i+1}(F_i) + \xi_i(t) =J(t),
\end{equation}
where
\begin{equation}
  \label{eq:shot_amplitude}
\langle \xi_i(t) \xi_j(t') \rangle = \frac{e}{A} \left[ \frac{ev^{(f)}(F_i)}{l}n_i + J^{-}_{i \to i+1}(F_i,n_{i + 1},T) + 2J^{-}_{i \to i+1}(F_i,n_i,T) \right] \delta_{ij} \delta(t - t').
\end{equation}
We see that $\eta(t)$ is independent of $i$, while $\xi_i(t)$ are independent Gaussian random variables associated with each site of the SSL. The strength of the fluctuations in the bias voltage may be tuned via the empirical parameter $\sigma_\eta$, while the strength of the fluctuations in the local tunneling current is completely determined by the parameters of Table~\ref{tab:params}.


\subsection{Disorder}
We also consider time-independent perturbations that break the periodicity of the SSL. We introduce variations in the widths of the wells and barriers via the scaling parameters $\beta_i$ and $\zeta_i$. The perturbed well and barrier lengths are
\begin{align}
\label{eq:disordered_lengths}
  w_i &= \beta_i w, \\
  d_i &= \zeta_i d.
\end{align}
The change in total length of the SSL modifies the bias constraint equation \eqref{eq:bias} to
\begin{equation}
V_{\text{bias}} = \sum_{i = 1}^{N} F_i l_i.
\end{equation}
To lowest order, the energy levels scale with $\beta_i$ according to
\begin{equation}
\varepsilon^{C,m}_{i} = \frac{\varepsilon^{C,m}}{\beta_{i}^2}.
\end{equation}
These modifications imply that the effective dielectric constant becomes dependent on $i$,
\begin{equation}
\varepsilon_i = l_i/(w_i/\varepsilon_w + d_i/\varepsilon_d).
\end{equation}
Following Bonilla et.~al.\ \cite{bonilla2006voltage}, equations~\eqref{eq:tunnel} and~\eqref{eq:backwards} are modified to account for the effects of disorder on $v^{(f)}(F_i)$ and $\tau_i$:
\begin{align}
  \label{eq:velocity_and_tau}
  v^{(f)}(F_i) &= \frac{\hbar^3}{2m^{*2}}\sum_{m = 1}^{3} \frac{ l_i \big(\gamma_{C1} + \gamma_{C,m}\big) \LARGE{\tau_i(\varepsilon^{i}_{C,m})}}{\big(\varepsilon_{i}^{C,1} - \varepsilon_{i + 1}^{C,m} + eF_i\big(d_i + \frac{w_i + w_{i + 1}}{2}\big) \big)^2 + \big(\gamma_{C,1} + \gamma_{C,m}\big)^2}\\
  \tau_i &= \frac{16k_i^2k_{i + 1}^2\alpha_i^2}{\big(k_i^2 + \alpha_i^2\big)\big(k_{i + 1}^2 + \alpha_i^2\big)\big(w_i + \frac{1}{\alpha_{i - 1}} + \frac{1}{\alpha_i}\big)\big(w_i + \frac{1}{\alpha_{i + 1}} + \frac{1}{\alpha_i}\big) e^{2\alpha_i d_i}}
\end{align}
The parameters $\gamma_{C,m}$ describe the width of the Lorentzian broadening functions that govern the degree to which the energy levels must be aligned in order for tunneling to take place. 
From reference \cite{bonilla2006voltage}, these are taken to be $\gamma_{C,1}=2.5$ meV, $\gamma_{C,2}=8.0$ meV, $\gamma_{C,3}=24$ meV.
The magnitudes of the propagating ($k_i^m$) or decaying ($\alpha_i^m$) wavevectors are given by
\begin{align}
\label{eq:k_tunnel}
\hbar k_i^m &= \sqrt{2m^{*}\varepsilon^{C,m}_i}\\
\hbar k_{i + 1}^m &= \sqrt{2m^{*}\Big(\varepsilon^{C,m}_{i} + e\left[d_i + \frac{1}{2}\big(w_i + w_{i + 1}\big)\right]F_{i}\Big)}
\end{align}
and
\begin{align}
\label{eq:a_tunnel}
  \hbar \alpha^m_{i - 1} &= \sqrt{2m^{*}\Big(e V_b + e\left[d_{i - 1} + \frac{w_i}{2}\right]F_i - \varepsilon^{C,m}_i\Big)}\\
  \hbar \alpha^m_i &= \sqrt{2m^{*}\Big(e V_b - e \left[ \frac{1}{2} w_i \right]F_i - \varepsilon^{C,m}_i \Big)}\\
  \hbar \alpha^m_{i + 1} &= \sqrt{2m^{*}\Big(e V_b - e \left[d_i + \frac{1}{2} w_i + w_{i + 1}\right]F_i - \varepsilon^{C,m}_i\Big)}
\end{align}

\section{Computing the Poincar\'e map}
\label{sec:method}

\begin{figure}
    \centering
    \includegraphics[scale=1]{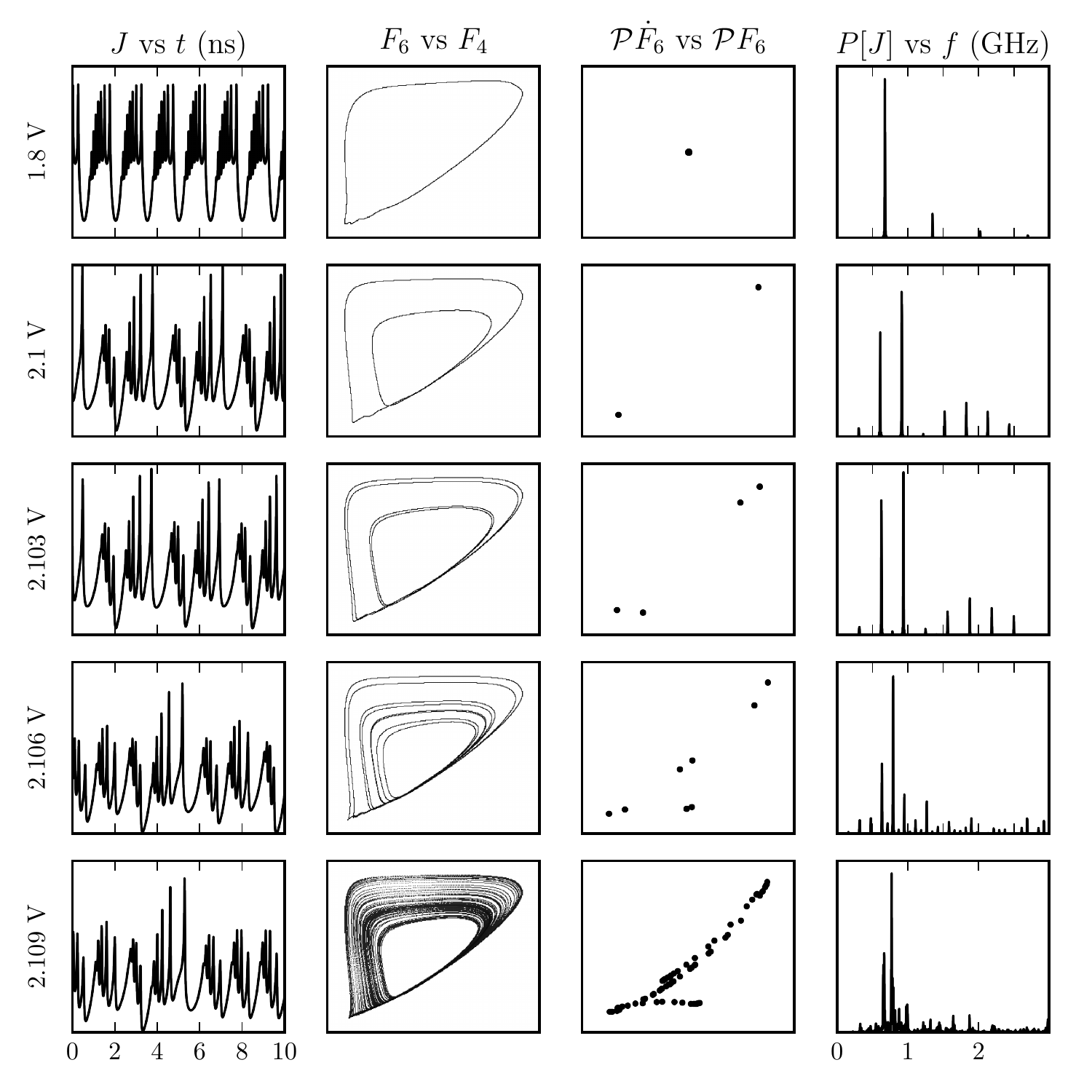}
    \caption{Representative phase portraits for the 10-period GaAs/Al$_{0.7}$Ga$_{0.3}$As SSL, taken from \cite{garcia2017}. The first column shows the average current $J$ plotted against time $t$. The second column shows the phase portrait $F_6(t)$ plotted against $F_4(t)$. The third column shows the Poincare map $\mathcal{P}\dot F_6(t^*)$ plotted against $\mathcal{P}F_6(t^*)$. The last column shows the power spectrum of $J(t)$. A periodic oscillation is shown in the first row. The period-doubling cascade to a chaotic attractor is shown in the bottom four rows.}
    \label{fig:portrait}
\end{figure}

The Poincar\'e map is used to gain insight into the structure of trajectories through high-dimensional space. In this section we outline our method of numerically computing the Poincar\'e map. The evolution equations~\eqref{eq:ST1}--\eqref{eq:STN} are evolved in time  using the forward Euler method and the trajectory $(F_i(t), n_i(t), J(t))$ through the $(2N+1)$-dimensional phase space is stored. When applicable, the stochastic terms $\xi_i(t)$ and $\eta(t)$ are included using the Euler-Maruyama method. 
The first step is to construct the phase portrait, i.e.\ to project the evolution onto a two-dimensional surface in phase space. We choose the surface spanned by the the coordinates $(F_i,F_j)$ for some values of $i$ and $j$ near the anode and cathode of the SSL.
Several phase portraits corresponding to a period-doubling cascade are illustrated in the second column of Figure~\ref{fig:portrait}.

\begin{figure}
    \centering
    \textsc{Power Spectra and Bifurcation Diagram}\\
    \includegraphics[scale=1]{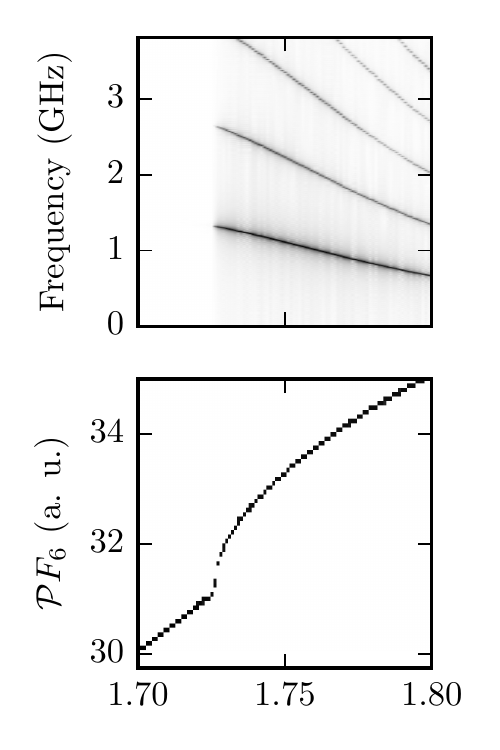}
    \hspace{-0.2in}
    \includegraphics[scale=1]{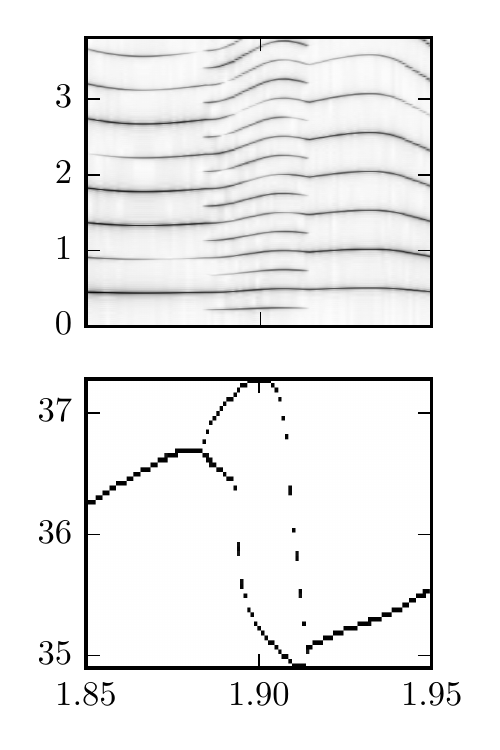}
    \hspace{-0.2in}
    \includegraphics[scale=1]{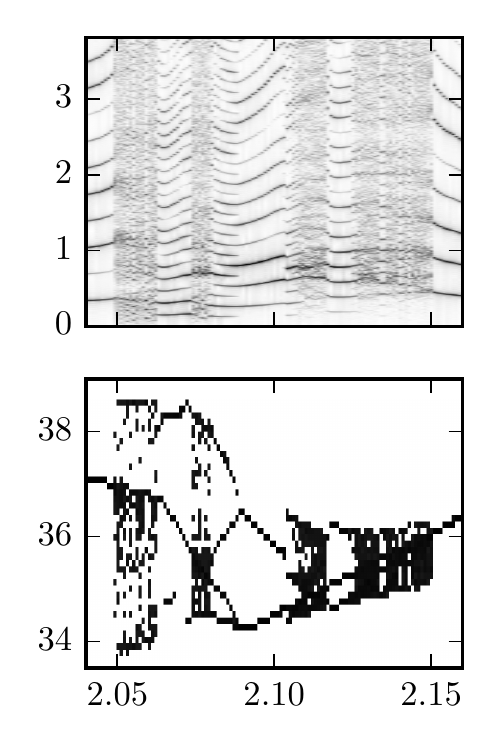}\\
    \vspace{-4ex} {\footnotesize Voltage (V)}
    \caption{ Power spectra and bifurcation diagram of a 10-period GaAs/Al$_{0.7}$Ga$_{0.3}$As SSL, taken from \cite{garcia2017}. (Top row) Power spectrum of $J(t)$ versus  voltage. (Bottom row) Bifurcation diagram of Poincar\'e map versus voltage. The Hopf bifurcation from the steady state is shown in the first column. A period doubling bubble is shown in the second column. A period-doubling cascade is shown in the third column.}
    \label{fig:bifurcation}
\end{figure}

The next step is to compute the Poincar\'e map of the phase portrait. After sufficient time has elapsed, and regardless of the initial conditions, the phase space trajectory settles onto one of the following time-invariant structures: (a) Fixed point, (b) periodic orbit, (c) strange attractor. The Poincar\'e map is used to distinguish between these structures. It is computed according to the following procedure: First, the transient behavior associated with the initial conditions is excised from the trajectory and only the remaining data is considered in what follows: When $F_i(t)$ passes through its median value, the time $t^*$ and the field $F_j(t^*)$ are stored. We also compute the quantity $\dot F_i(t^*)$ from equation~\eqref{eq:ST1}. We then discard all of the values of $t^*$ for which $\dot F_i(t^*)>0$, in order to prevent sampling the same orbit more than once per cycle. The remaining points constitute the Poincar\'e map $\mathcal{P}F_i$. The Poincar\'e map transforms the essentially continuous trajectory through phase space into a discrete map from the one-dimensional interval onto itself \cite{collet2009iterated}. We represent it visually in terms of (a) phase portraits, plotting $\mathcal{P}F_j(t^*)$ against $\mathcal{P}\dot F_j(t^*)$, as in the third column of Figure~\ref{fig:portrait}, or (b) bifurcation diagrams, plotting $\mathcal{P}F_j$ against some external parameter such as the bias voltage, as in the bottom row of Figure~\ref{fig:bifurcation}.

Both fixed points and periodic orbits appear as a single point in the visualization of the Poincar\'e map. However, fixed points are easily distinguished from periodic orbits (one-cycles) by computing the power spectrum associated with the current $J(t)$:
\begin{equation}
\label{eq:fourier_transform}
	P[J](f) = \left|\int_{t_i}^{t_f}dt\, e^{-i 2\pi ft} J(t)\right|^2,
\end{equation}
where $f$ is the frequency.
A period-doubling bifurcation is identified when  one-cycles transition to two-cycles, producing two points in the Poincar\'e map.
Chaotic regions are identified where a proliferation of period-doubling bifurcations occur, and the number of points in the Poincar\'e map increases without bound, yielding a fractal structure in the bifurcation diagram.
Dynamical structures revealed by the Poincar\'e map are associated with various power spectra as follows: (a) Periodic orbits correspond to a series of peaks with widths of the same order as the frequency bin size, falling at integer multiples of the fundamental frequency, (b) period doubling bifurcations are recognized when the number of peaks in the spectrum changes by a factor of two, and a new peak appears in the power spectrum at half the fundamental frequency, (c) chaotic attractors exhibit power spectra containing both sharp peaks and broadband noise. We illustrate the bifurcation diagram and power spectrum in Figure~\ref{fig:bifurcation}.

\section{Results}
\label{sec:results}


Dynamical instabilities are found in two distinct \emph{plateaus}, over which the local electric fields of the SSL cease to increase monotonically as a function of $V_{\text{bias}}$. The \emph{first plateau} occurs at very low voltages, where tunneling transport between the ground states of adjacent wells are nearly aligned with one another in energy. The \emph{second plateau} occurs in the region of $V_{\text{bias}}$ such that the external electric field tilts the potential of the SSL to align the ground state of well $i$ with first excited state of well $i+1$. We do not observe a third plateau because the third excited state becomes unbound at bias voltages that align it with the ground state of the previous well.

Within a plateau, we may observe period-doubling bifurcations, period-doubling cascades, and chaotic attractors, whose locations depend upon on the values of the rest of the parameters, in particular $\sigma$, the contact conductivity, $N$, the number of wells making up the SSL, and $\beta$ ($\zeta$), the strength of the perturbations to the well (barrier) widths. Shorter superlattices exhibit faster oscillations and a greater variety of dynamical behavior in the second plateau \cite{garcia2017}. We are concerned with finding the parameter regions with the strongest nonlinear phenomena. Our observable of interest is the periodicity, i.e. the number of distinct points in the Poincar\'e map, which is equivalent to the number of branches in the bifurcation diagram. The nonlinear orbits of higher periodicity are found deeper into the period doubling cascade, either within or near to the chaotic windows.



\begin{figure}[ht]
	\centering
	\subfloat[\tall{} SSL \label{fig:sigmaa}]{%
		\includegraphics[scale=0.75]{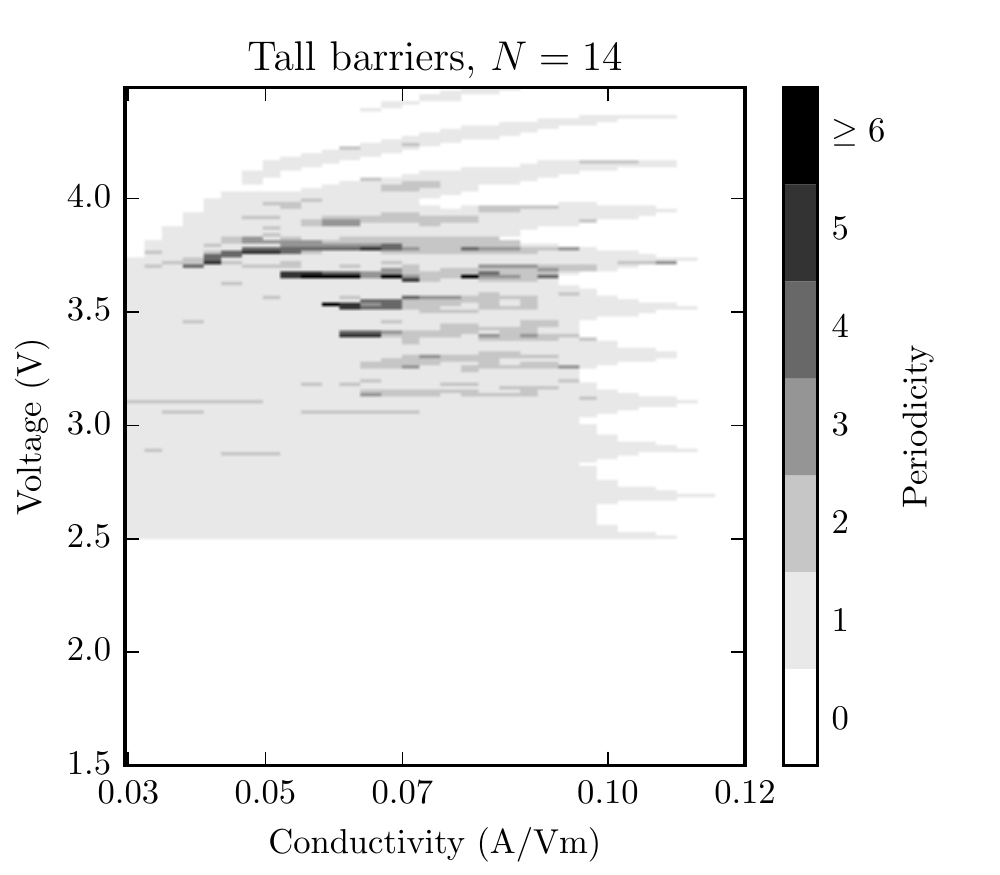}}
	\subfloat[\short{} SSL \label{fig:sigmab}]{%
		\includegraphics[scale=0.75]{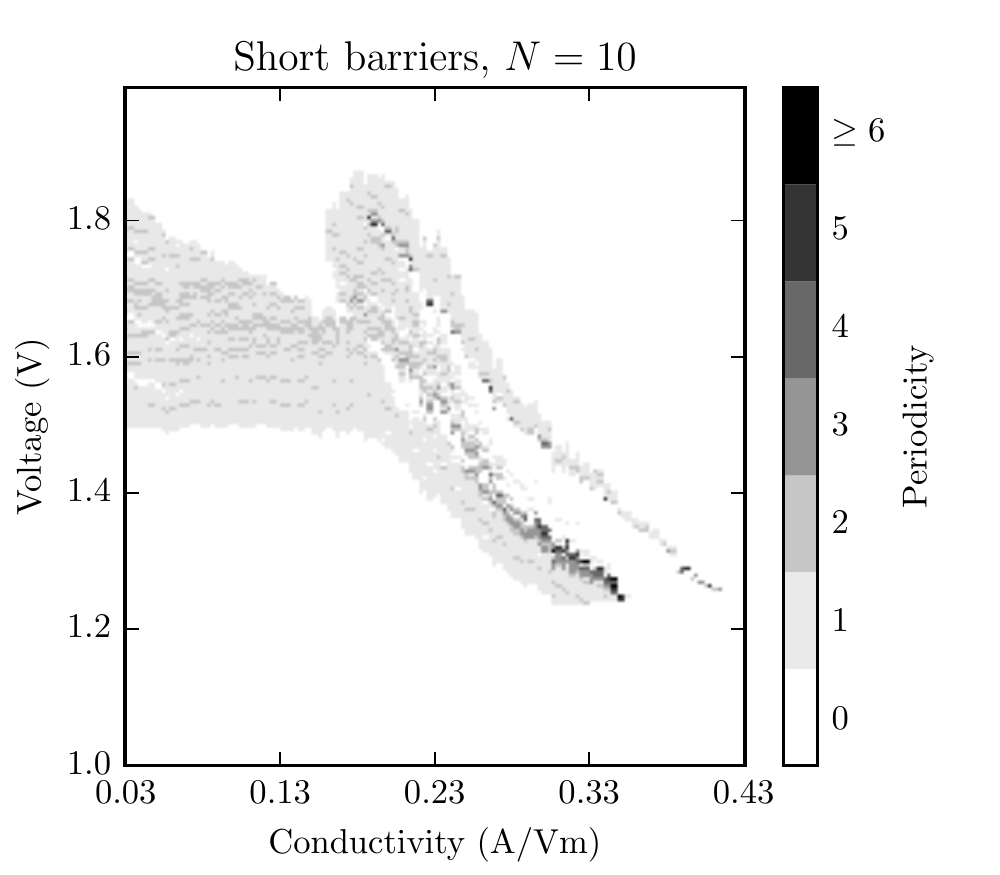}}
	\caption{Varying the conductivity. The second plateau is shown in the figures. It exists for low conductivities, then fragments and disappears for higher conductivity. The black dots and bands indicate orbits of high periodicity.}
	\label{fig:sigma}
\end{figure}
An important empirical parameter of the SRT model is the contact conductivity. In Figure~\ref{fig:sigma}, we show the response of the periodicity to variations in the contact conductivity. We have chosen the values of $N$ which maximize the total area of the high-period orbits in the Poincar\'e mapping as a function of voltage: $N=14$ for \tall{} SSLs, and $N=10$ for \short{} SSLs.

In both cases we observe that the second plateau remains in existence for very low conductivity, then narrows, fragments, and disappears at sufficiently high conductivity.
The highest-period orbits and chaos are concentrated at high conductivity and the highest voltages contained in the second plateau. These results suggest that in order to find the most chaotic dynamics, the highest possible conductivities that allow for the existence of the second plateau should be sought out. In the rest of our results, we set the conductivity to $\sigma=0.06$ A/Vm for \tall{} SSLs (the same as in \cite{garcia2017}) and $\sigma=0.30$ V/Am for \short{} SSLs.

\begin{figure}[ht]
	\centering
	\subfloat[\tall{} SSL \label{fig:Na}]{
		\includegraphics[scale=0.75]{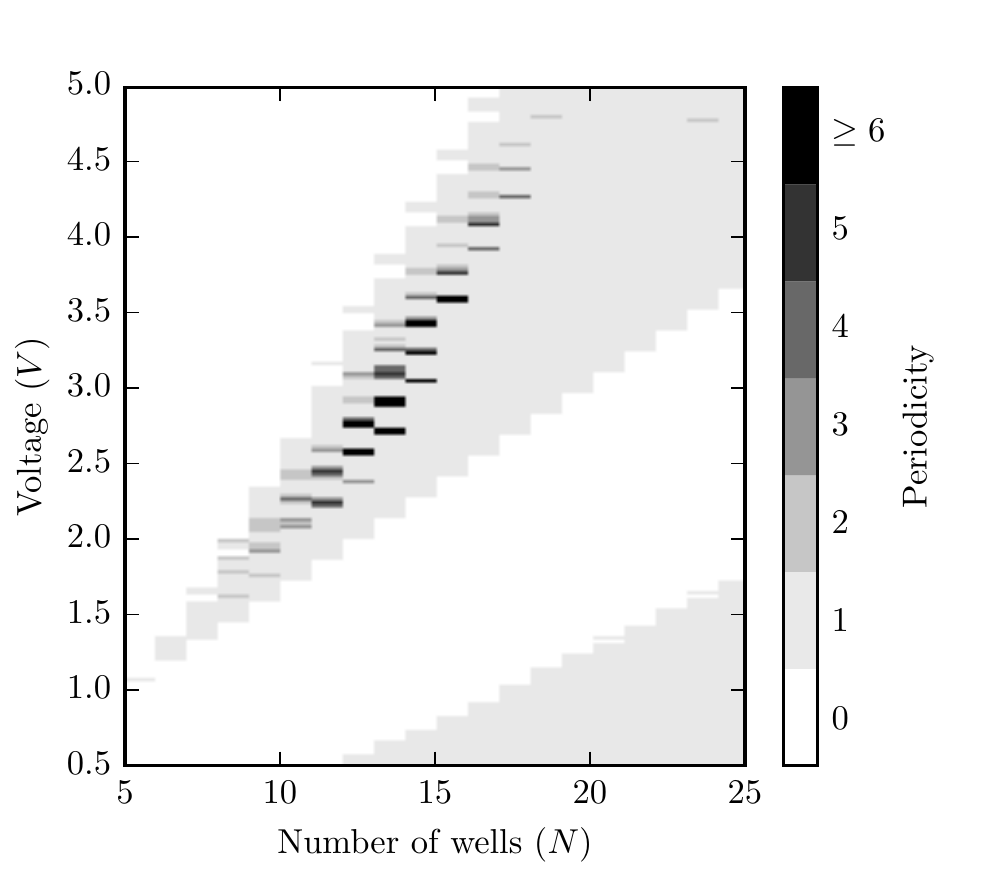}}
	\subfloat[\short{} SSL \label{fig:Nb}]{%
		\includegraphics[scale=0.75]{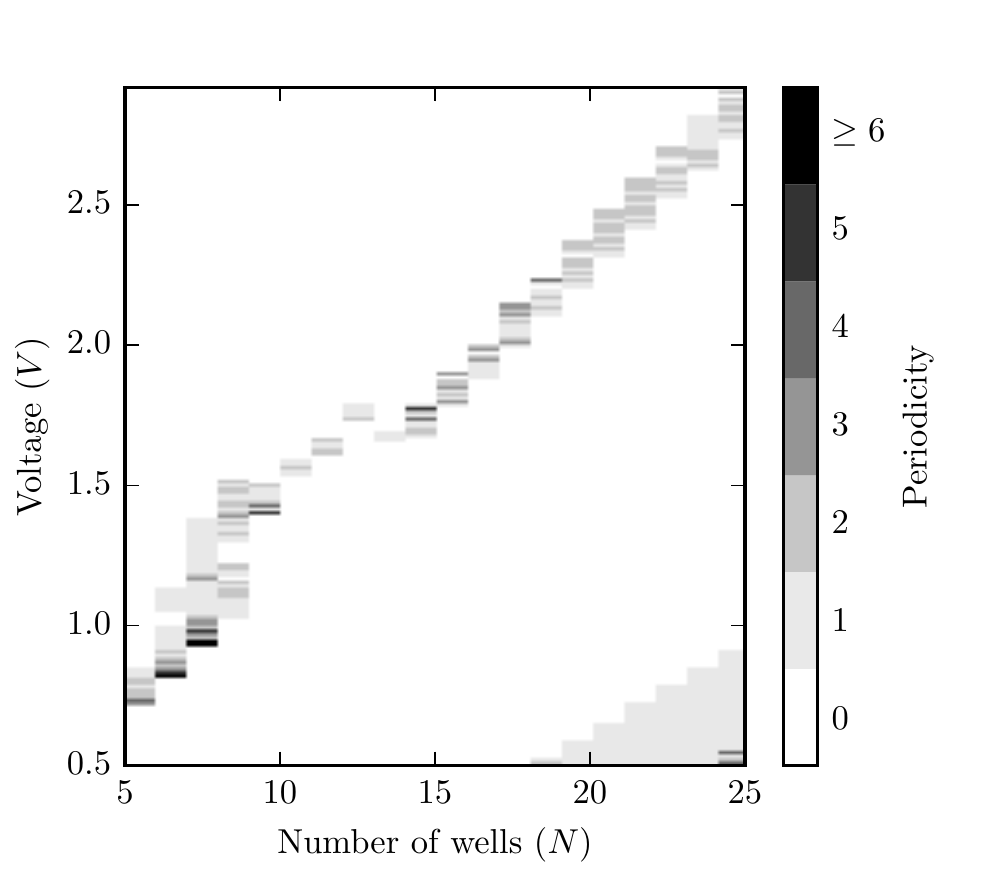}}
	\caption{Varying the number of wells $N$ in the superlattice. A band of higher period orbits appears around
	$N=10$, both for the tall barriers on the left and shorter barriers on the right. For the latter barriers the windows narrow to become hardly observable between $N=11$ and $N=14$.}
	\label{fig:N}
\end{figure}

We next consider varying $N$, the number of periods making up the SSL in Figure~\ref{fig:N}. In both cases, we observe a band of higher periodicity including chaotic behavior in the vicinity of $N=10$. The shorter superlattice appears to have a gap in the chaotic behavior between N=10 and N=15, but in fact the voltage windows containing the chaos are too narrow to be observed at this scale.
The band of chaotic behavior in the \tall{} SSLs is located along the higher voltages of the second plateau. In both cases, the chaotic windows narrow and finally close for values of $N$ between 15 and 20. Part of the first plateau is also visible in the bottom right corners of Figures~\ref{fig:Nb}(a) and (b).

\begin{figure}[ht]
	\centering
	\subfloat[\tall{} SSL \label{fig:betaa}]{%
		\includegraphics[scale=0.75]{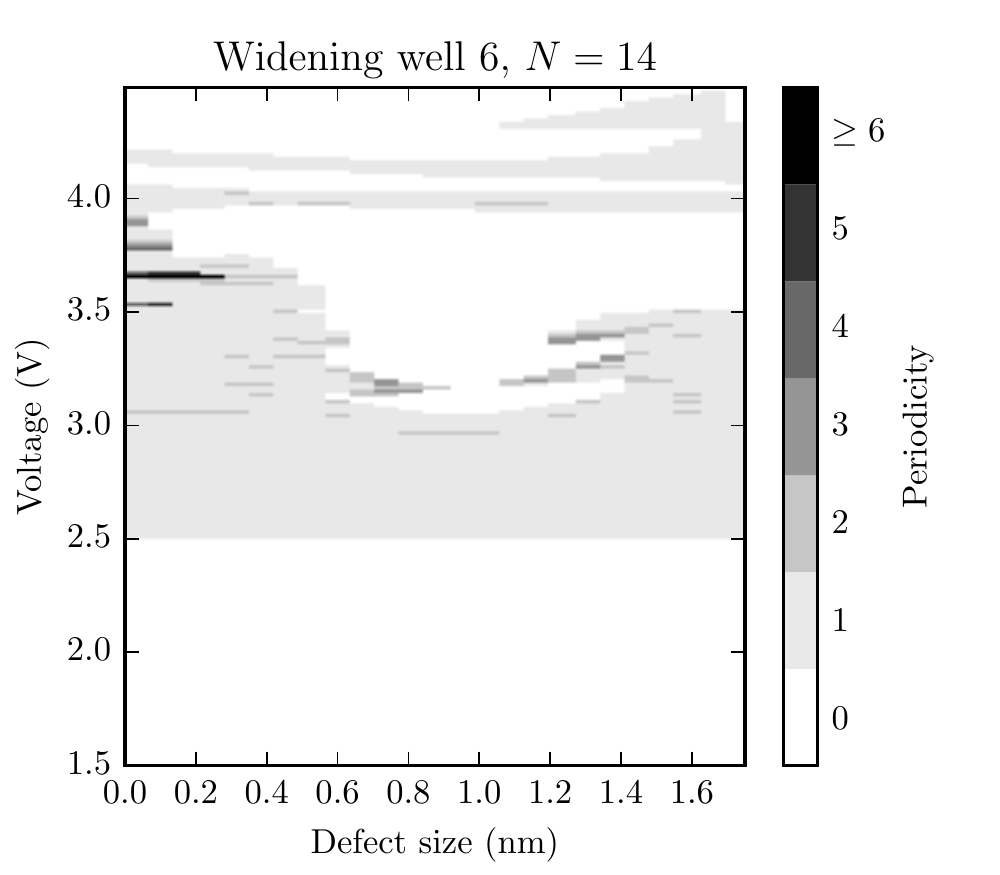}}
	\subfloat[\short{} SSL \label{fig:betab}]{%
		\includegraphics[scale=0.75]{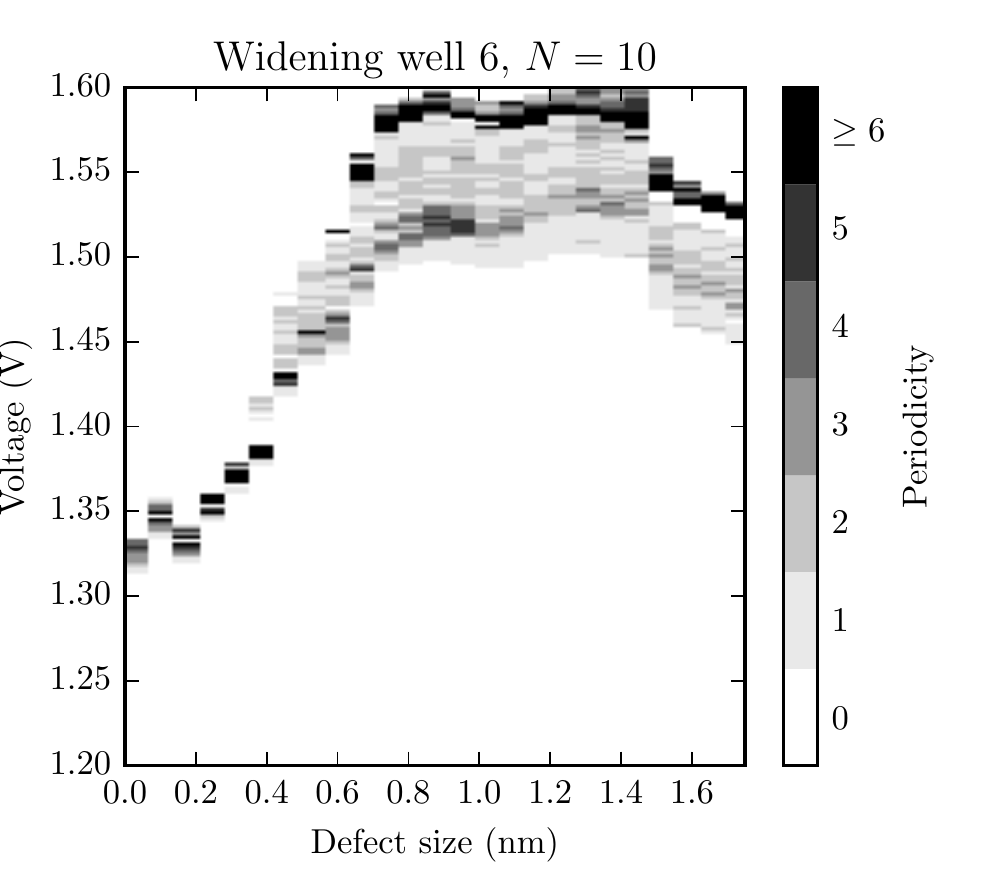}}
	\caption{Varying the disorder $\beta$. On one hand, the tall barriers on the left permit high periodicity only for small values of $\beta$, but these features are destroyed for higher values of disorder. On the other hand, the short barriers on the right show greatly enhanced high periodicity at relatively large values ($\beta\geq0.65$~nm) of disorder.}
	\label{fig:beta}
\end{figure}
Finally, we investigate the sensitivity of the chaos to disorder by varying $\beta$ in Figure~\ref{fig:beta}. The width of one GaAs monolayer is about 0.28~nm. In our simulations, the addition of a single monolayer is capable of destroying the chaos in the case of the taller barriers. On the other hand, the chaotic signal of the shorter barriers appears to be enhanced by the presence of disorder. We note that the location of the added disorder is nearer to the cathode for the taller barriers and nearer to the anode for the shorter barriers. It would would be interesting to further investigate the conditions where the chaos is enhanced by the presence of disorder. 

\begin{figure}[ht]
	\centering
	\includegraphics[scale=1]{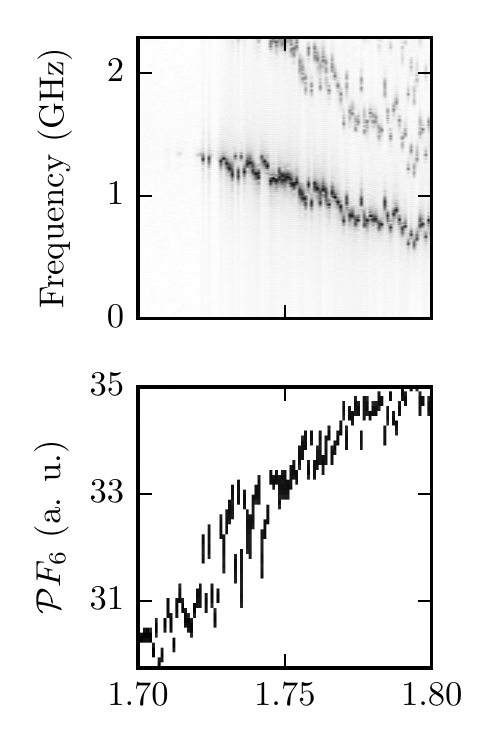}
	\hspace{-0.2in}
	\includegraphics[scale=1]{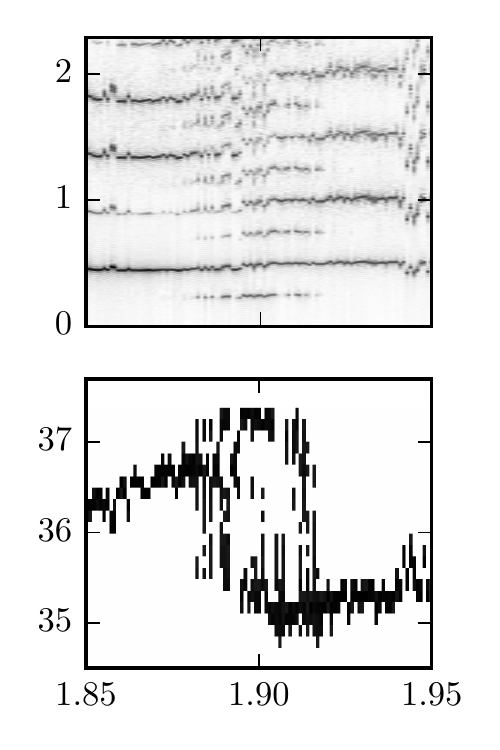}
    \hspace{-0.2in}
	\includegraphics[scale=1]{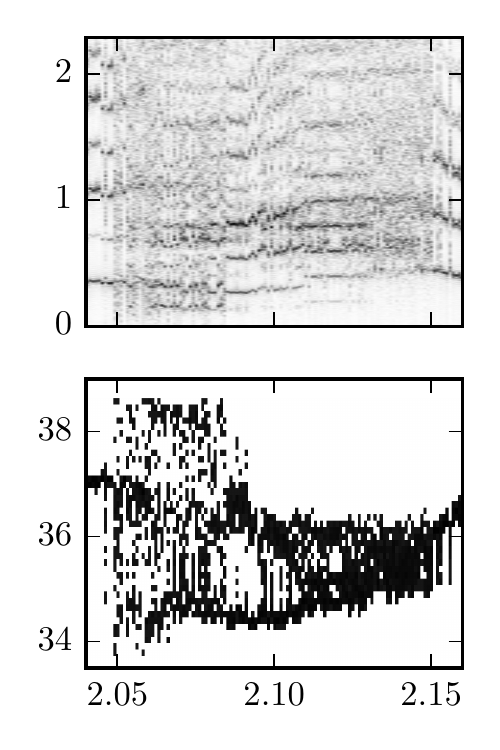}\\
	\vspace{-4ex} {\footnotesize Voltage (V)}
	\caption{Including shot noise and bias noise for the N=10 \tall{} SSL; the simulation parameters are otherwise the same as in Figure~\ref{fig:bifurcation}. The addition of noise widens the branches of the bifurcation diagram obscures the fine details of the period-doubling cascade.}
	\label{fig:noise}
\end{figure}
We simulate the effects of noise on the bifurcation diagram in Figure~\ref{fig:noise}, including both shot noise and bias voltage noise. We can see that in regions where the periodicity (number of branches in the bifurcation diagram) is low, the noise widens the Poincar\'e map from an isolated point into a cluster, but the branches are still recognizable. On the other hand, where the periodicity is higher or the dynamics are chaotic, the noise widens the Poincar\'e map into a broad band.

Let us imagine an experiment which detects the local field $F_6$, shown in Figure~\ref{fig:noise}, at some finite resolution, i.e.\ the number of bins, with the objective of reading out a sequence of random bits. Then the random bit generation rate will scale proportionally to the width of the Poincar\'e map times the resolution of the imaginary $F_6$-sensor. The bands within the regions of higher periodicity or intrinsic dynamical chaos would cover a larger number of bins. Hence these regions would generate random bits at higher bandwidth in comparison with the regions of lower periodicity. In practice, the local current $I_{6}$ would be easier to measure, but the results would be qualitatively very similar.


\section{Conclusions}
\label{sec:conclusion}
The discovery of robust, high-frequency, intrinsic nonlinear phenomena and chaos in shorter semiconductor superlattices in the sequential tunneling regime points the way toward a variety of useful devices. In this work, we have characterized the response of the chaotic oscillations to variations in the number of SSL periods and the contact conductivity, providing a guide for the experimental investigation of the emergence of chaos in short SSLs. The chaos is predicted to appear as the result of a period-doubling cascade.

We have also investigated the response of the chaotic signal to stochastic perturbations in the local tunneling currents and the bias voltage. In contrast with the slower, noise-driven chaos in the first plateau for longer superlattices, we observe that shorter SSLs allow for faster, intrinsic chaos in the second plateau.

We have also investigated the effects of variations in the widths of the wells and barriers on the period-doubling cascade. We find that the period-doubling cascade is very sensitive to these perturbations. An error of only a single monolayer has a strong impact on the width of the windows of chaotic behavior. The chaotic windows may be either widened or suppressed depending on the location of the irregularities, hence it may be possible to engineer aperiodicities in SSLs in order to increase the chaotic signal.

We had initially conjectured that the presence of aperiodicities could unfold the period-doubling bifurcation into a second Hopf bifurcation. However, this turned out not to be the case. Our study of the DC-biased SRT model shows only a period-doubling route to chaos (no second Hopf bifurcation). On the experimental side, quasi-periodic orbits and the associated invariant tori are commonplace. An interesting theoretical question is: By what mechanism do quasi-periodic orbits appear in weakly-coupled SSLs?

\FloatBarrier

\acknowledgments
 This material is based upon work supported by, or in part by, the U. S. Army Research
Laboratory and the U. S. Army Research Office under contract/grant number
 444045-22682 and by the Ministerio de Econom\'\i a y Competitividad
of Spain under grant MTM2014-56948-C2-2-P. MRG also acknowledges support from MECD through the FPU program.


\begin{thebibliography}{99}
\bibitem{sti95} D.R. Stinson, \emph{Cryptography: Theory and Practice, 3rd ed.} (CRC Press, Boca Raton, 2006).

\bibitem{gal08} R.G. Gallager, \emph{Principles of Digital Communication} (Cambridge University Press, Cambridge, UK, 2008).

\bibitem{nie00} M.A. Nielsen, I.L. Chuang, \emph{Quantum Computation and Quantum Information} (Cambridge University Press, Cambridge, UK, 2000).

\bibitem{asm07} S. Asmussen, P.W. Glynn, \emph{Stochastic Simulation: Algorithms and Analysis} (Springer-Verlag, New York, 2007).

\bibitem{Leo92} Karl Leo and Peter Haring Bolivar and Frank Br\"uggemann and Ralf Schwedler and Klaus K\"ohler. Observation of Bloch oscillations in a semiconductor superlattice. Solid State Communications {\bf10}, 943--946 (1992).

\bibitem{Ka97} Kastrup, J and Hey, R and Ploog, KH and Grahn, HT and Bonilla, LL and Kindelan, M and Moscoso, M and Wacker, A and Gal{\'a}n, J. Electrically tunable GHz oscillations in doped GaAs-AlAs superlattices. Phys Rev B. {\bf 55}, 2476 (1997).

\bibitem{Wu99} Wu, JQ and Jiang, DS and Sun, BQ. Room-temperature microwave oscillation in AlAs/GaAs superlattices. Physica E: Low-dimensional Systems and Nanostructures, 137--141 (1999).

\bibitem{uch08} A. Uchida, K. Amano, M. Inoue, K. Hirano, S. Naito, H. Someya, I. Oowada, T. Kurashige, M. Shiki, S. Yoshimori, K. Yoshimura, P. Davis, Fast physical random bit generation with chaotic semiconductor lasers. Nat. Photonics {\bf 2}, 728-732 (2008).

\bibitem{murphy08} T. E. Murphy and R. Roy,  The world's fastest dice. Nat Photonics. {\bf 2}, 714-715 (2008)

\bibitem{reid09} I. Reidler,  Y. Aviad,  M. Rosenbluh, I. Kanter, Ultrahigh-speed random number generation based on a chaotic semiconductor laser. Phys Rev Lett. {\bf 103}, 024102 (2009)

\bibitem{kanter10} I. Kanter, Y. Aviad,  I. Reidler, E. Cohen, M. Rosenbluth. An optical ultrafast random bit generator. Nat Photonics. {\bf 4}, 58 (2010).

\bibitem{scia15} M. Sciamanna, K.A. Shore, Physics and applications of laser diode chaos. Nature Photonics {\bf 9}, 151-162 (2015).

\bibitem{li2013fast} W. Li, I. Reidler, Y. Aviad, Y. Y. Huang, H. Song, Y. H. Zhang, M. Rosenbluh, and I. Kanter, Fast Physical Random-Number Generation Based on Room-Temperature Chaotic Oscillations in Weakly Coupled Superlattices,
Phys. Rev. Lett. \textbf{111}, 044102 (2013)

\bibitem{bonilla2005nonlinear} L. L. Bonilla and H. T. Grahn, Non-linear dynamics of semiconductor superlattices. Reports on Progress in Physics {\bf 68}, 577 (2005)

\bibitem{alvaro2014noise} M. Alvaro, M. Carretero, and L. Bonilla, Noise-enhanced spontaneous chaos in semiconductor superlattices at room temperature.  EPL (Europhysics Letters)  {\bf 107}, 37002 (2014)

\bibitem{garcia2017} Ruiz-Garcia, M. and Essen, J. and Carretero, M. and Bonilla, L. L. and Birnir, B., Enhancing chaotic behavior at room temperature in GaAs/(Al,Ga)As superlattices, Phys. Rev. B {\bf 95}, 085204 (2017).

\bibitem{Gal96a} B. Galdrikian and B. Birnir, Period Doubling and Strange Attractors in Quantum Wells. Phys. Rev. Lett. {\bf 76}, 3308 (1996)

\bibitem{Bat03} A. A. Batista, B. Birnir, P. I. Tamborenea and D. S. Citrin, Period-doubling and Hopf bifurcations in far-infrared driven quantum well intersubband transitions. Phys. Rev. B {\bf 68}, 035307 (2003)

\bibitem{ama02} A. Amann, J. Schlesner, A. Wacker, E. Sch\"oll, Chaotic front dynamics in semiconductor superlattices. Phys. Rev. B {\bf 65}, 193313 (2002).

\bibitem{huang2012experimental} Y. Huang, W. Li, W. Ma, H. Qin, and Y. Zhang, Experimental observation of spontaneous chaotic current oscillations in GaAs/Al0.45Ga0.55As superlattices at room temperature.  Chinese Science Bulletin  {\bf 57}, 2070 (2012)

\bibitem{huang2013spontaneous} Y. Huang, W. Li, W. Ma, H. Qin, H. T. Grahn, and Y. Zhang, Spontaneous quasi-periodic current self-oscillations in a weakly coupled GaAs/(Al,Ga)As superlattice at room temperature.  Applied Physics Letters  {\bf 102}, 242107 (2013)

\bibitem{bonilla2002} L. L. Bonilla, Theory of Nonlinear Charge Transport, Wave Propagation and Self-oscillations in Semiconductor Superlattices. Journal of Physics: Condensed Matter {\bf 14}, R341 (2002)

\bibitem{bonilla2009nonlinear} L. L. Bonilla and S. W. Teitsworth, \emph{Nonlinear wave methods for charge transport}. (Wiley VCH, Weinheim, 2009)

\bibitem{bonilla2000microscopic} L. L. Bonilla, G. Platero, and D. S\'anchez, Microscopic derivation of transport coefficients and boundary conditions in discrete drift-diffusion models of weakly coupled superlattices.  Phys. Rev. B  {\bf 62}, 2786 (2000)

\bibitem{bon2016jmi} L.L. Bonilla, M. Alvaro, and M. Carretero, Chaos-based true random number generators. Journal of Mathematics in Industry {\bf 7}, 1 (2016).

\bibitem{Yin16} Z. Yin, Y. Zhang, M. Ruiz-Garc\'{i}a, M. Carretero, L. L. Bonilla, K. Biermann and H. T. Grahn, Noise-enhanced chaos in a weakly coupled GaAs/(Al,Ga)As superlattice, Phys. Rev. E {\bf 95}, 012218.

\bibitem{collet2009iterated} P. Collet and J. Eckmann, \emph{Iterated Maps on the Interval as Dynamical Systems}, Modern Birkh{\"a}user Classics (Birkh{\"a}user Boston, 2009).

\bibitem{Gr87} R. Graham, Squeezing and frequency changes in harmonic oscillations. Journal of Modern Optics  {\bf 34}, 873 (1987)












\bibitem{Bat02} A. A. Batista, P. I. Tamborenea, B. Birnir, M. Sherwin, and D. S. Citrin, Nonlinear dynamics in far-infrared driven quantum-well intersubband transitions. Phys. Rev. B {\bf 66}, 195325 (2002)

\bibitem{wacker2002semiconductor} A. Wacker, Semiconductor superlattices: a model system for nonlinear transport. Phys. Rep. {\bf 357}, 1--111 (2002)

\bibitem{luo1999controllable} Luo K J, Teitsworth S W, Kostial H, Grahn H T and Ohtani N, Controllable bistabilities and bifurcations
in a photoexcited GaAs/AlAs superlattice. Appl. Phys. Lett. {\bf 74} 3845 (1999)





\bibitem{bonilla2006voltage} L. L. Bonilla, R. Escobedo, and G. Dell'Acqua, Voltage switching and domain relocation in semiconductor superlattices.  Phys. Rev. B  {\bf 73}, 115341 (2006)

\bibitem{guido2006hopf} G. Dell'Acqua, L. L. Bonilla, and R. Escobedo, Hopf Bifurcation in a Superlattice Model, Proceedings of the International Conference on Computational and Mathematical Methods in Science and Engineering (2006).

\end{thebibliography}
\end{document}